\documentclass[universe,review,accept,pdflatex,moreauthors]{mdpi} 
\firstpage{1} 
\pubvolume{9}
\issuenum{4}
\articlenumber{178}
\pubyear{2023}
\copyrightyear{2023}
\externaleditor{Academic Editor: {Roman Pasechnik} %MDPI: please add academic author name
}
\datereceived{13 February 2023} 
\daterevised{30 March 2023} % Comment out if no revised date
\dateaccepted{31 March 2023} 
\datepublished{4 April 2023} 
\usepackage{amssymb}% http://ctan.org/pkg/amssymb
\usepackage[nohyperlinks,printonlyused]{acronym}
\usepackage{multicol}
\usepackage{pifont}% http://ctan.org/pkg/pifont
\newcommand{\cmark}{\ding{51}}%
\newcommand{\xmark}{\ding{55}}%

\definecolor{mygreen}{rgb}{.1,.75,.1}

\hreflink{https://doi.org/10.3390/\linebreak universe9040178} % If needed use \linebreak
\Title{{A Concise} %
 Review on Some Higgs-Related New Physics Models in Light of Current Experiments}

\TitleCitation{A Concise Review on Some Higgs-Related New Physics Models in Light of Current Experiments}

\Author{{Lei Wang} %MDPI: Please carefully check the accuracy of names and affiliations.
 $^{1,\dag}$\orcidA{}, {Jin Min Yang} %MDPI: we corrrect the name to right format. please confirm 
 $^{2,3,\dag}$\orcidB{}, Yang Zhang $^{2,4,\dag}$\orcidC{}, Pengxuan Zhu $^{2,\dag}$\orcidD{} and Rui Zhu $^{2,3,}$*$^{,\dag}$\orcidE{}}

\AuthorNames{Lei Wang, Jin Min Yang, Yang Zhang, Pengxuan Zhu and Rui Zhu}

\AuthorCitation{Wang, L.;{ Yang, J.M.}; %Authors: we corrected the author's name.
Zhang, Y.; Zhu, P.; Zhu, R.}
\address{%
$^{1}$ \quad Department of Physics, Yantai University, Yantai 264005, China; leiwang@ytu.edu.cn\\
$^{2}$ \quad CAS Key Laboratory of Theoretical Physics, Institute of Theoretical Physics,
                Chinese Academy of Sciences, Beijing 100190, China; jmyang@itp.ac.cn {(J.M.Y.)}; %Authors:we corrected the author's name.
                zhangyangphy@zzu.edu.cn (Y.Z.); \linebreak{}zhupx99@icloud.com (P.Z.)\\
$^{3}$ \quad School of Physical Sciences, University of Chinese Academy of Sciences,
                Beijing 100049, China \\
$^{4}$ \quad School of Physics, Zhengzhou University, Zhengzhou 450000, China
                }

\corres{Correspondence: zhurui@itp.ac.cn}

\firstnote{{These authors} %MDPI: please add sepcial symbol to the authors
 contributed equally to this work.}
\abstract{The Higgs boson may serve as a portal to new physics beyond the standard
model (BSM), which is implied by the theoretical naturalness or experimental anomalies. This review aims to briefly survey some typical Higgs-related BSM models. First, for the theories to solve the hierarchy problem, the two exemplary theories, the low energy supersymmetry (focusing on the minimal supersymmetric model) and the little Higgs theory, are discussed. For the phenomenological models without addressing the hierarchy problem, we choose 
the two-Higgs-doublet models (2HDMs) to emphatically elucidate their phenomenological power in explaining current measurements of muon $g-2$, the $W$-boson mass and the dark matter (DM) data. 
For the singlet extensions, which are motivated by the cosmic phase transition and the DM issue,  we illustrate the singlet-extended standard model (xSM) and the singlet-extended 2HDM (2HDM+S),
emphasizing the vacuum stability. In the decade since the discovery of the Higgs boson, these theories have remained the typical candidates of new physics, which will be intensively studied in future theoretical and \mbox{experimental research.}}

\keyword{Higgs boson; beyond standard model (BSM) physics; minimal supersymmetric standard model (MSSM)} 

\begin{document}

%%%%%%%%%%%%%%%%%%%%%%%%%%%%%%%%%%%%%%%%%%
\section{Introduction}

\begin{figure}[H]
	%\centering
	\includegraphics[width=0.6\linewidth]{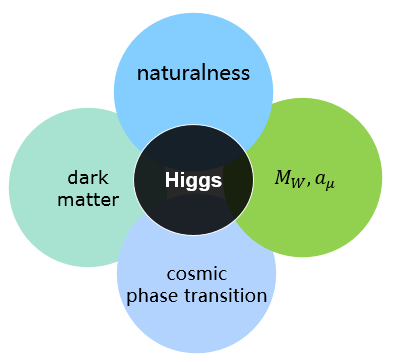} 
	\caption{A sketch map showing that the Higgs sector may serve as a portal to BSM physics implied by the naturalness,  the experimental anomalies from the muon $g-2$ or $W$-mass, the cosmic phase transition and DM.\label{fig1}}
\end{figure}

With the discovery of the 125 GeV Higgs boson~\cite{ATLAS:2012yve, CMS:2012qbp},
high energy physics has entered the post-Higgs era, in which the main goal
is to test the Higgs properties and explore new physics BSM.
Despite the fact that the current experiments such as the LHC
and the DM direct detections have found no clear evidence of new particles,
our belief in the existence of BSM physics has never been shaken.
This is because the SM is obviously not the ultimate theory due to the problems such as
the naturalness, the vacuum stability, the {neutrino} %MDPI: we change the footnote to note. please confirm
%Response: we have confirmed.
 mass\endnote{Neutrinos in the SM (the active neutrinos) are massless. However, the explanation of the neutrino mass by introducing right-handed neutrino $N_R$ via the Yukawa interaction like other fermion fields might not be the whole story of nature. Since $N_R$ is sterile, the gauge symmetry allows $N_R$ to acquire the Majorana mass $M$, and therefore it does not pair up with the active neutrino to make up a Dirac fermion. $N_R$ carries the lepton number, so the neutrino mass is often related to the flavor physics. If $M$ is very large, the only dimension-5 operator allowed by the SM symmetries can generate the active neutrino mass of order $v^2/M$, where $v$ is the SM Higgs vev. This idea is called the ``seesaw'' mechanism. The neutrino masses may be also closely related to the origin of flavor mixings,  the CP violation and the fermion mass hierarchy, and the neutrino phenomenology is relatively far from the Higgs field. So, in this review, we will not discuss neutrinos further. For the reviews on neutrinos, see, e.g., ~\cite{Fritzsch:1999ee, Xing:2003ez, Mohapatra:2006gs, Altarelli:2004za}.}, the DM and the matter--antimatter asymmetry in the
universe. All these problems seem to be caused or related to the Higgs sector.
In other words, the Higgs sector may serve as a portal to the BSM physics implied by
theoretical naturalness or experimental anomalies (such as muon $g-2$ and $W$-mass)
or cosmic observations (such as DM and matter--antimatter asymmetry), as illustrated 
in Figure~\ref{fig1}.

From the theoretical side, it is well-known that the observed mass of the Higgs boson leads to a naturalness problem in the SM.
Obtaining a Higgs mass of 125 GeV requires an extreme fine-tuning of the model parameters.
So from the theoretical point of view, the BSM physics should
solve the quadratic divergence of the Higgs boson \cite{Arkani-Hamed:2012fhg, Kawamura:2013xwa, Feng:2013pwa}. In this end, the low energy SUSY is the most popular paradigm (for a comprehensive review, see, e.g.,~\cite{Haber:1984rc}, while
for recent brief reviews, see, e.g., \cite{Baer:2020kwz, Wang:2022rfd, Yang:2022qyz}).   
In addition, the quadratic divergence of the Higgs boson mass can be canceled at the one-loop level 
in the little Higgs theory \cite{Arkani-Hamed:2001nha, Arkani-Hamed:2002iiv}, 
while in the theories with large \cite{Arkani-Hamed:1998jmv} or warped extra dimensions 
\cite{Randall:1999ee}, the naturalness can be obtained by reducing the fundamental scale to a weak scale.  

From the experimental side, the $W$-boson mass recently measured by the CDF II deviates from the SM 
by $7\sigma$~\cite{CDF:2022hxs}, while the muon $g-2$ measured by FNAL and BNL deviates from the SM by $4.2\sigma$~\cite{Muong-2:2021ojo}. Note that the CDF result disagrees with the recent LHC measurement~\cite{LHCb:2021bjt, ParticleDataGroup:2022pth}, which is in agreement with the SM prediction, albeit with a relatively large uncertainty. For the muon $g-2$ anomaly, the lattice calculations \cite{Colangelo:2022vok,Ce:2022kxy,Borsanyi:2020mff,FermilabLattice:2022izv,Alexandrou:2022amy} 
seem to shift up the SM value to relax the deviation from $4.2\sigma$ to $1.5\sigma$, showing a $2.1\sigma$ tension with the $e^+e^-$ data-driven determination of the HVP contribution.
The BSM physics should be able to jointly explain both anomalies plus the DM. 
So far, some models have been found to be feasible, such as the low energy SUSY~\cite{Yang:2022gvz,Domingo:2022pde,Tang:2022pxh} 
and some specific 2HDMs~\cite{Han:2022juu,Babu:2022pdn,Kim:2022hvh,Arcadi:2022lpp,Chen:2023eof}
(for recent brief reviews, see, e.g., \cite{Wang:2022yhm}).

\vspace{-10pt} {}

\textls[-30]{From the cosmic side, a Higgs field may be a portal to the cosmic cold DM (for reviews see, e.g., \cite{Arcadi:2019lka}, and for recent studies see, e.g., \cite{Kim:2023pwf}) and may also trigger the electroweak phase transition in the early universe. The stability of the current electroweak vacuum and the cosmological phase transition can be studied from the Higgs potential. In explaining the baryon asymmetry of the universe, a FOPT is required in the electroweak baryogenesis to provide the departure from the thermal equilibrium (for reviews of baryogenesis, see, e.g., \cite{Cohen:1993nk, Dine:2003ax, Cline:2006ts, Morrissey:2012db, Elor:2022hpa}). However, it is well known that the EWPT from the SM Higgs is a smooth crossover, i.e., the SM cannot produce a FOPT. In the new physics models such as the singlet extensions, the newly introduced particles and interactions may change the Higgs potential, giving a FOPT and inducing the detectable gravitational waves. On the other hand, the DM, either scalar or fermion, may exist in some hidden sector that couples to the visible sector very weakly via the Higgs portal, and the scalar potential in such a hidden sector may also trigger a FOPT and induce the gravitational~waves.} 

In this note, we briefly survey some typical Higgs-related BSM physics models, including the low energy SUSY (focusing on the minimal SUSY model), the little Higgs models, the 2HDMs and the simple singlet extensions of the Higgs sector. For each illustrated BSM model, we will emphatically discuss its phenomenological power in light of current measurements of the muon $g-2$, the $W$-boson mass and the DM. For the singlet extensions, we will emphasize the induced cosmic phase transition and the DM relic density as well as the vacuum stability. The demonstrated numerical results are from our previous works, whereas we try to cite relevant works as completely as possible.   

\section{\label{sec:2}  Low Energy SUSY}

\subsection{A Light Higgs Boson in SUSY} 
So far, the LHC experiments are consistent with the elementary Higgs boson predicted by SM.
To accommodate such a light elementary scalar particle, low energy SUSY is the most natural 
framework \cite{witten2021}.

In the SM, the masses of fermions or gauge bosons are prohibited by gauge or chiral symmetry. However, the Higgs boson mass is not protected by any symmetry, and it has a quadratic divergence from loop corrections. Therefore, it is sensitive to the UV cut-off energy scale.
However, in SUSY, the quadratic divergences from the loop corrections to the Higgs boson mass are ``technically'' canceled out and only logarithmic divergences remain. So, the Higgs boson mass is stabilized at the weak scale, which is not sensitive to the UV cut-off energy scale.

\begin{figure}[H]
	%\centering
	\includegraphics[width=0.98\linewidth]{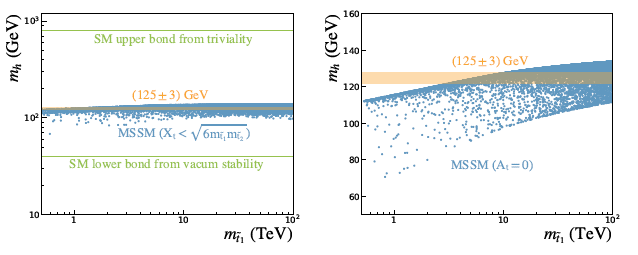} 
	\caption{\label{mh-mass} The scatter plots showing the mass of the SM-like Higgs boson
versus the stop mass in the MSSM. The SM upper and lower bounds are from the requirement of vacuum
stability and non-triviality for the UV cut-off scale around {TeV} %MDPI: Please ensure that permission has been obtained and there is no copyright issue. If copyright is needed, please provide a citation in the following format: “Reprinted/adapted with permission from Ref. [XX]. Copyright year, copyright owner’s name”. More details on “Copyright and Licensing” are available via the following link: https://www.mdpi.com/ethics#10.
%Response: we ensure that there is no copyright issue.
 \cite{Riesselmann:1997kg}.
} 
\end{figure}

In SUSY, due to the holomorphicity requirement of the Yukawa couplings, the Higgs sector must be extended to two Higgs doublets $H_u$ and $H_d$ with opposite hypercharges to give masses to both up-type and down-type quarks after electroweak symmetry breaking. As the most economical SUSY model, the MSSM predicts five Higgs bosons, among which the lightest CP-even $h$ is the SM-like Higgs in the decoupling limit, i.e., all other Higgs bosons being sufficiently heavier than the $Z$-boson mass $m_Z$, the couplings of $h$ with the SM particles approach the SM predictions. $m_h$ is upper bounded by about 135 GeV in the~MSSM, 
\begin{equation}
	m_{h}^2 \sim m_Z^2 \cos^2{2\beta} + \frac{3 m_t^4}{2\pi^2 v^2} \left[\log{\frac{M_S^2}{m_t^2}} + \frac{X_t^2}{M_S^2}\left(1 - \frac{X_t^2}{12 M_S^2}\right) \right], 
\end{equation}
where $v =\sqrt{ v_u^2 + v_d^2} = 246~{\rm GeV}$ with the vevs of two Higgs fields $v_u\equiv \langle H_u \rangle$ and $v_d \equiv \langle H_d \rangle$, $\beta$ is defined by $\tan{\beta} = v_u / v_d$, $M_S$ is the geometric average of two stop masses $M_S = \sqrt{m_{\tilde{t}_1} m_{\tilde{t}_2}}$ defined to be the SUSY-breaking scale, and $X_t$ is the stop mixing parameter given by $X_t = A_t - \mu /\tan{\beta}$ with $A_t$ being the stop soft trilinear coupling and $\mu$ being {the} %Authors: we changed "a" to "the".
 higgsino mass parameter.  
Therefore, a larger value of $\tan{\beta}$ is required to maximize the tree-level contribution $m_Z |\cos{2\beta}|$, a large $M_S$ value is favored to enhance the logarithmic contribution, and a large stop trilinear coupling $X_t = \sqrt{6}M_S$ can enhance the stop loop contribution \cite{Carena:2013qia}. 
To see this clearly, we perform a scan using the package \textsc{FeynHiggs}~\cite{Heinemeyer:1998yj}, where the higher-order corrections are from the two-loop level and from the log-resummations at the NNLL level. We vary the sensitive parameters in \mbox{the ranges}
\begin{equation}\begin{split}
		0.5{\rm ~TeV} \leq M_{Q3}=M_{U3}=M_{D3} \leq 100{\rm ~TeV},
		\quad \vert X_{t} \vert \leq \sqrt{6 m_{\tilde t_1}  m_{\tilde t_2}}, 
		\quad 1 < \tan{\beta} < 50. 
\end{split}
\end{equation}
where $M_{Q3}$, $M_{U3}$, and $M_{D3}$ are the third-generation squark soft masses. Higgsinos and gauginos are assumed to not be so heavy, i.e., $M_1=M_2/2=1$ TeV and 
\begin{equation}
100{\rm ~GeV} \leq \mu \leq 350{\rm ~GeV}.
\end{equation}
{Other} %MDPI: please confirm if keep no indent format. same for below
%Response: yes, here should keep no indent format.
 soft mass parameters, such as the gluino mass $M_3$, are fixed at 100 TeV.  We see from Figure~\ref{mh-mass} that for a stop in the range of 0.5--100 TeV, the mass of $h$ is approximately in the range of 80--135 GeV. This means that a light Higgs boson is predicted in the MSSM. In comparison, in the SM, the Higgs mass is a free parameter.  The requirement of vacuum stability \cite{Altarelli:1994rb,Casas:1994qy} and non-triviality \cite{Hambye:1996wb}  restrain the Higgs mass in the range of 40--800 GeV if the UV cut-off scale is around TeV \cite{Riesselmann:1997kg}.

Therefore, the LHC discovery of a light Higgs boson around 125 GeV can be regarded as a triumph of low energy SUSY. On the other hand, from the detailed analysis in \cite{Cao:2012fz} or our Figure~\ref{mh-mass} above, we see that for a moderate $ A_t $ in magnitude, the 125 GeV SM-like Higgs boson mass requires a relatively heavy stop above several hundred GeV, which is consistent with the lower mass bound around 500 GeV from the null search results of stops at the LHC \cite{susy-moriond-2021} (note that the LHC bounds on the plane of stop mass versus the LSP mass show that for a sizable mass splitting between stop and LSP, the lower bound on the stop mass is about 1.2 TeV while for a stop mass near the LSP mass plus the top quark mass, the lower bound on the stop mass is only about 500 GeV). For a very small size or zero value of  $A_t$, we see from  Figure~\ref{mh-mass} that the 125 GeV SM-like Higgs boson mass requires a stop mass heavier than 3 TeV.

\subsection{DM, Muon {$g-2$} %MDPI: please confirm this word format. g-2 or $g-2$, please keep same fromat in whole tex
%Response: we changed every case of g-2 to $g-2$ and of W to $W$ in this manuscript for consistency.
 and {$W$}-Mass in SUSY}  
\begin{figure}[H]
\begin{adjustwidth}{-\extralength}{0cm}\centering
	\includegraphics[width=0.87\linewidth]{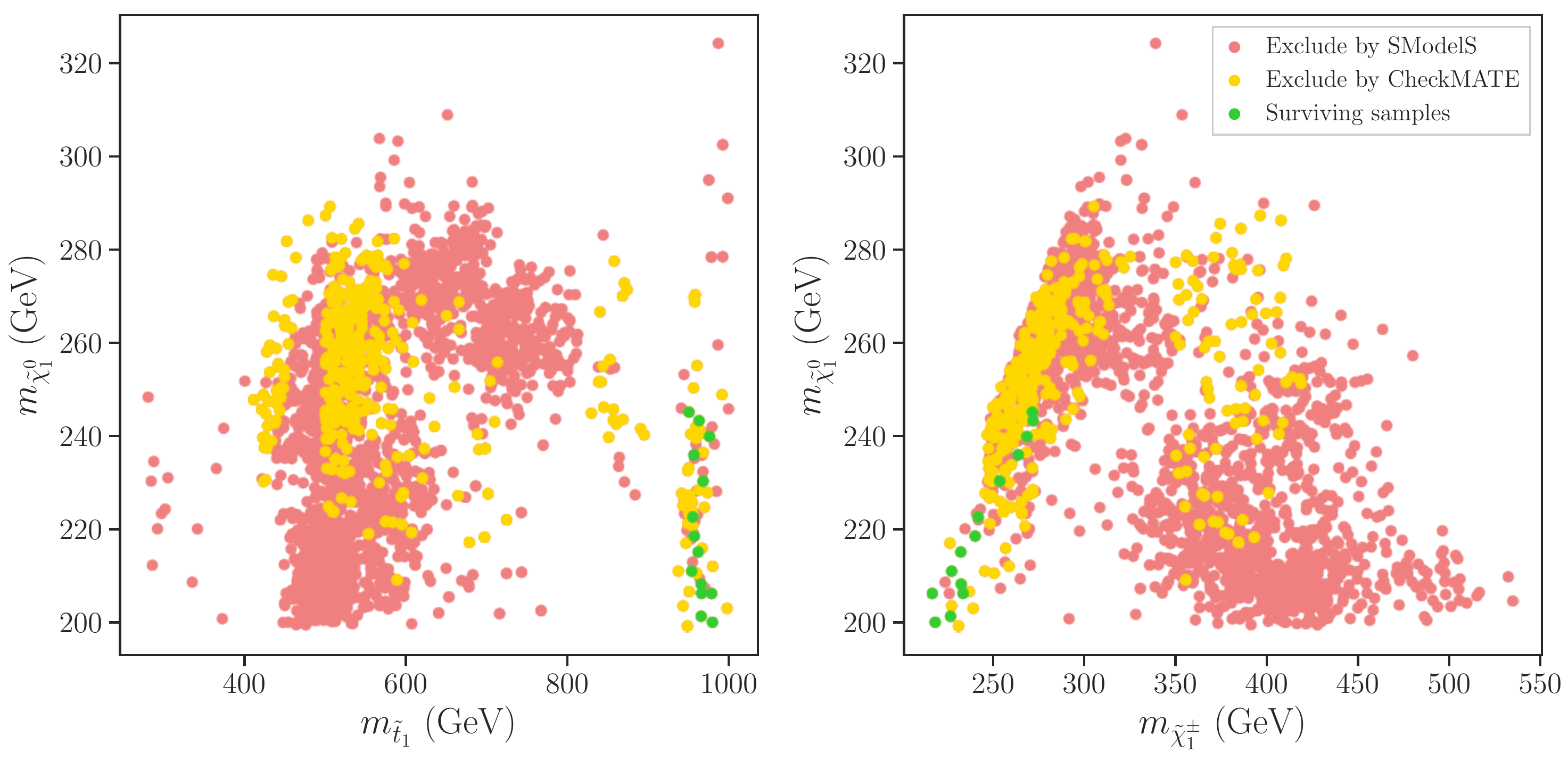} 
\end{adjustwidth}
	\caption{\label{mw-susy}The scatter plots jointly explaining at $2\sigma$ level 
the muon $g-2$ reported by the FNAL and the $W$-boson mass measured by the CDF II as well as 
providing the correct DM relic density under direct detection limits. 
This figure is taken from our previous {work} %MDPI: Please ensure that permission has been obtained and there is no copyright issue. If copyright is needed, please provide a citation in the following format: “Reprinted/adapted with permission from Ref. [XX]. Copyright year, copyright owner’s name”. More details on “Copyright and Licensing” are available via the following link: https://www.mdpi.com/ethics#10.  The same for below.
%Response: we ensure that there is no copyright issue and the same for below.
 \cite{Yang:2022gvz}.}
\end{figure}

In addition to naturally predicting the Higgs boson mass, the beauty of SUSY is also reflected in its ability to jointly explain the muon $g-2$ reported by the FNAL and the $W$-boson mass measured by the CDF II as well as provide the observed DM relic density under direct detection limits \cite{Yang:2022gvz,Domingo:2022pde}. 
However, such a joint explanation requires a light stop just below 1 TeV in the MSSM~\cite{Yang:2022gvz}, as shown in Figure~\ref{mw-susy}, which should be accessible at the next run of the LHC.   

\begin{figure}[H]
\begin{adjustwidth}{-\extralength}{0cm}
	\centering
	\includegraphics[width=0.48\linewidth]{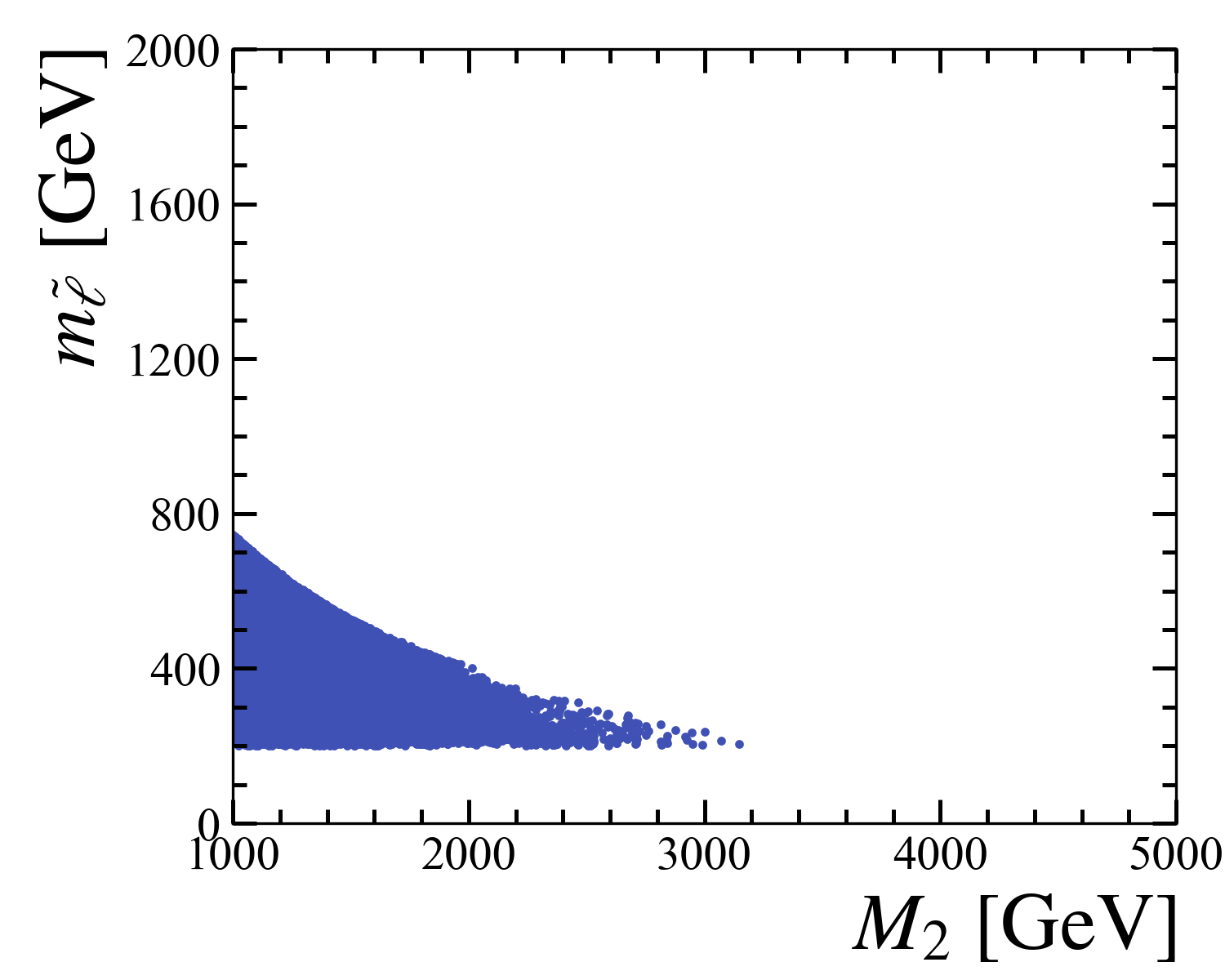} 
	\includegraphics[width=0.48\linewidth]{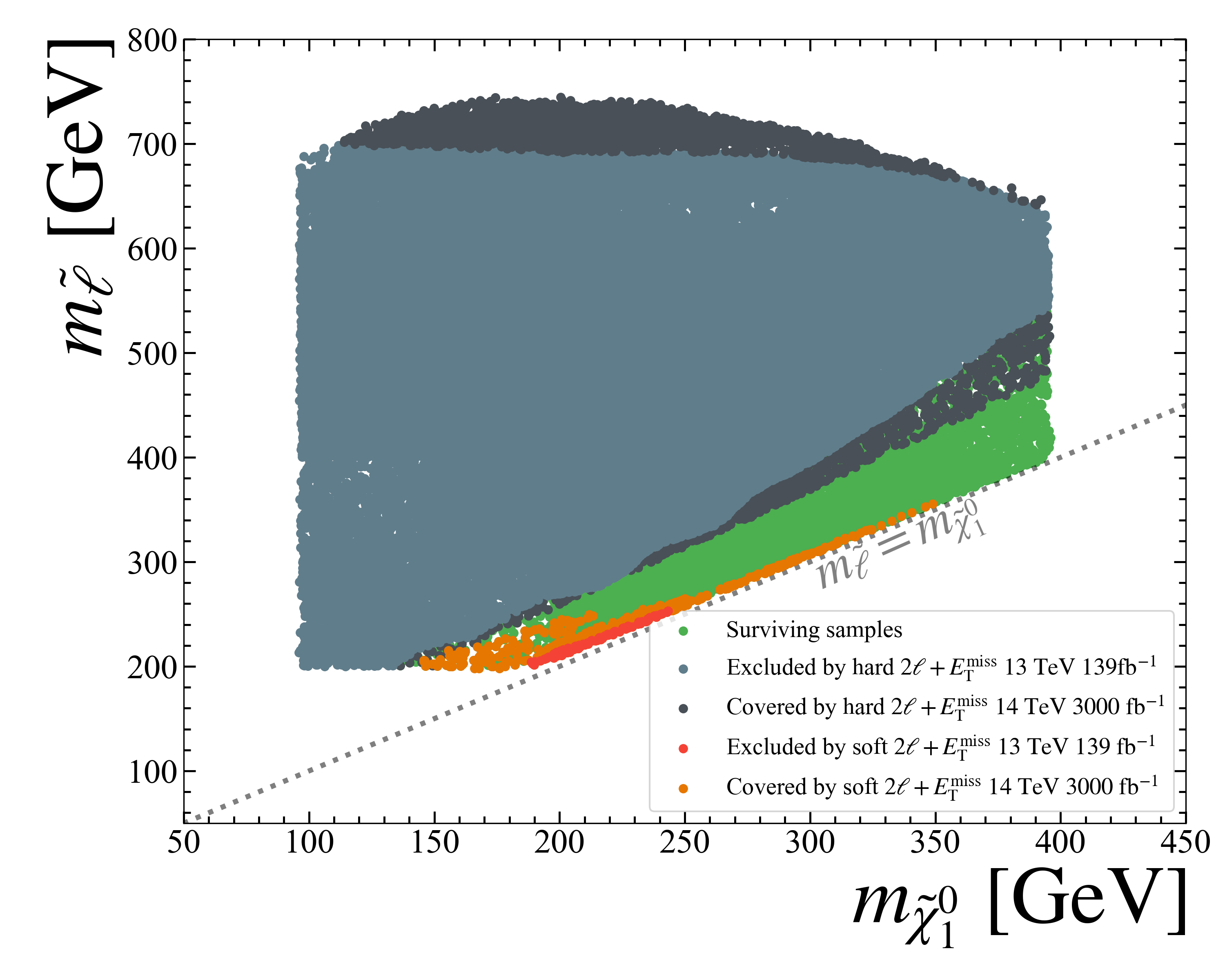} 
\end{adjustwidth}
	\caption{\label{gm2-mssm}\textls[-15]{The scatter plots explaining at $2\sigma$ level 
the muon $g-2$ reported by the FNAL, with the higgsino ($\mu$ in the range of 100--400 GeV) 
as the lightest super particle satisfying the  $2\sigma$ upper bound of the DM relic density 
and the direct detection limits.  This figure is taken from our previous {work}
 \cite{Zhao:2022pnv}.}}
\end{figure}

Note that without the anomaly of the $W$-boson mass, 
the single anomaly of the muon $g-2$ can be readily explained in various low energy effective SUSY models 
\cite{Chakraborti:2020vjp, Chakraborti:2021kkr, Chakraborti:2021dli, Chakraborti:2021squ, Chakraborti:2021ynm, Chakraborti:2021mbr, Chakraborti:2022sbj, Abdughani:2019wai, Cox:2018qyi, Athron:2021iuf, Wang:2021bcx, Ning:2017dng, Abdughani:2021pdc, Cao:2021tuh, Wang:2021lwi, Cao:2022htd, Cao:2018rix, Cao:2022chy, Cao:2022ovk, Zhao:2021eaa, Yang:2021duj, Zhang:2021gun, Cao:2019evo, Cao:2021lmj, Wang:2022wdy, Li:2021poy, Barman:2022jdg, Zhao:2022pnv, Ajaib:2023jhc}.
In any case, the sleptons in the loop contributions to the muon $g-2$ cannot be too heavy and 
may be discovered at the HL-LHC, as shown in Figure~\ref{gm2-mssm}. 

\vspace{-10pt}

One of the most appealing features of low energy SUSY is that the lightest neutralino can act as a typical WIMP, which remains one of the most extensively studied particles despite not having been experimentally probed yet. At the sub-TeV scale, the current direct detection data on DM favor the WIMP being a gauge singlet particle, such as the singlino in the NMSSM\endnote{In the representation of $SU(2)_L$, the MSSM can be seen as a realization of the ``minimal DM models''. In this view, the higgsino (wino) DM in SUSY is a typical triplet (doublet) DM in the minimal DM model \cite{Cirelli:2005uq, Cirelli:2009uv, DiLuzio:2018jwd}.}. Hence, the muon $g-2$ and the $W$-boson mass anomaly would have more significance in the phenomenology of LHC/HL-LHC. On the other hand, the higgsino DM that exceeds TeV, for instance, is considerably beyond the limits of LHC probing.
 
\begin{figure}[H]
 	%\centering
 	\includegraphics[width=0.57\linewidth]{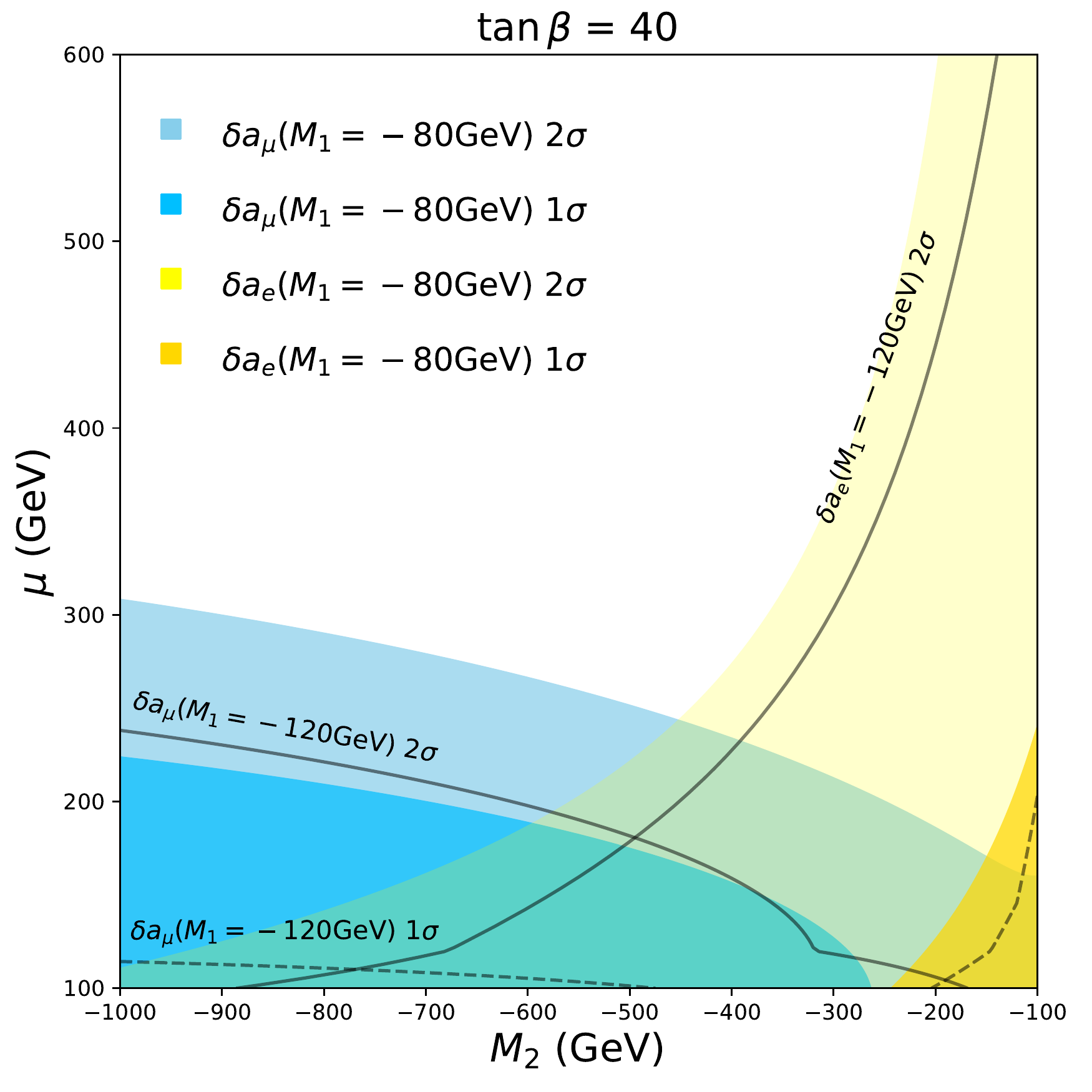} 	
 	\caption{\label{gm2-both}{The} %MDPI: we move the figure after its first citation, and re-order the whole reference. please confirm
  %Response: we have confirmed.
 MSSM parameter space explaining both anomalies of 
 		the muon $g-2$ reported by the FNAL and the electron $g-2$ from the Berkeley experiment~\cite{Parker:2018vye}.
 		This figure is taken from our previous work \cite{Li:2021koa}.
 	}
 \end{figure}

What is unusual\endnote{The combined explanation often requires the introduction of a flavor violation, see Ref.~\cite{Crivellin:2018qmi}.} is that the MSSM may be the common source of \cite{Li:2021koa,Li:2022zap} the muon\linebreak $g-2$ anomaly and the negative $2.4\sigma$ deviation of the electron $g-2$ between the experimental value~\cite{Hanneke:2008tm} and the SM prediction~\cite{Aoyama:2019ryr} from the measurement of the fine structure constant by the Berkeley experiment~\cite{Parker:2018vye}.
As shown in Figure~\ref{gm2-both}, such a joint explanation at the $2\sigma$ level requires a specific parameter space, in which a rather light bino-like LSP and a light higgsino-like NLSP as well as a large $\tan\beta$ are used to predict a positive contribution to the muon $g-2$ and a negative contribution to the electron $g-2$.
In this case, the thermal freeze-out of the bino-like LSP  gives an over-abundance and cannot be assumed to be the DM candidate (the DM can be some superWIMP as light as GeV and produced from the late-decay of the thermally freeze-out bino-like LSP). Note that the MSSM is found \cite{Carena:2012np,Carena:1997ki,Menon:2009mz} to be unable to realize a FOPT due to the current lower mass bound on the stop mass. 

\textls[-15]{However, the MSSM with boundary conditions at the cut-off
scale, say the CMSSM or mSUGRA with boundary conditions at the GUT scale, cannot 
explain the muon $g-2$ or the $W$-boson mass because sleptons and stops are both too
 heavy~\cite{Wang:2021bcx}. To explain the muon $g-2$, the boundary conditions at the cut-off scale have to be relaxed~\cite{Akula:2013ioa, Wang:2015rli, Wang:2018vrr, Li:2021pnt}. }

\par The NMSSM~\cite{Ellwanger:2009dp} extends the MSSM with a gauge singlet superfield, and thus it contains one extra neutralino, named a singlino. With the continuously improving sensitivity of DM direct detections, the singlino-like DM candidate~\cite{Cao:2018rix, Cao:2019qng} has some special phenomenological advantages. For example, the parameter space of the higgsino-like DM in MSSM can not provide a sufficient relic density in explaining the $g-2$ anomaly in front of the LHC data; while in the NMSSM, the corresponding parameter space can be released for the singlino-like DM co-annihilating with the higgsino-like NLSP to relax the experimental tensions. Moreover, the electroweakinos may generate sufficiently large contributions to the $W$-boson mass~\cite{Domingo:2022pde}.

The extra singlet Higgs can also achieve the non-zero vev in electroweak symmetry breaking, which can predict a FOPT \cite{Pietroni:1992in, Davies:1996qn, Huber:2006wf, Huang:2014ifa, Kozaczuk:2014kva, Athron:2019teq, Baum:2020vfl, Borah:2023zsb} in single- or multi-steps and may also induce the possibly detectable gravitational waves \cite{Wang:2022lxn, Borah:2023zsb}.

In summary, so far the low energy SUSY can survive all current experiments and explain the 
plausible experimental anomalies. The light relevant sparticles required by the explanations of 
anomalies, e.g., light sleptons required by the muon $g-2$  \cite{Abdughani:2019wai,Zhao:2022pnv} 
or light stops required by the CDF II $W$-mass  \cite{Yang:2022gvz}, should be accessible at the ongoing
LHC or the forthcoming HL-LHC. The only problem of low energy SUSY is that it now has a little fine-tuning 
(percent level for the phenomenological MSSM and per-mille level for the CMSSM \cite{Han:2016gvr}) 
caused by sizably heavier stops than the weak scale\endnote{However, an analysis \cite{vanBeekveld:2019tqp} gave a ten percent level fine-tuning for low-energy SUSY. }. 
To tackle the little fine-tuning caused by heavy stops, the idea of supersoft stops was proposed to screen 
the UV-sensitive logs \cite{Cohen:2020ohi}.

\section{\label{sec:3}  Little Higgs Models}
\subsection{A Light Higgs in Little Higgs Models } 
Another popular way to obtain a light Higgs boson naturally is to make the Higgs boson a pseudo-Nambu--Goldstone boson of some broken global symmetry. This idea was proposed in the 1970s,
 and based on this idea, a model was constructed \cite{Kaplan:1983fs,Kaplan:1983sm,Georgi:1984af} 
in the 1980s. However, this still needs fine-tuning from the weak scale to the cut-off scale. 
In the early 21st century, inspired by dimension (de)construction \cite{Arkani-Hamed:2001kyx}, 
the collective symmetry breaking mechanism was introduced to build 
some little Higgs models \cite{Arkani-Hamed:2001nha,Arkani-Hamed:2002iiv}. 
 The little Higgs is usually classified into the composite Higgs models, which are not surveyed here (for a review, see, e.g., \cite{Bellazzini:2014yua}).  
  
The collective symmetry breaking mechanism is the key point to give the Higgs boson a small mass
without incurring a quadratic divergence at the one-loop level. 
In this mechanism, at least two kinds of interactions (say gauge interactions with coupling constants 
$g_1$ and $g_2$) 
are introduced and they collectively work to break the global symmetry to give the Higgs boson a mass.
Switching off any of them, i.e.,  $g_1=1$ or $g_2=0$, the remaining interaction is not sufficient to fulfill
this mission. 
In realization, different groups and breaking modes can be selected.
According to how the electroweak gauge group of the SM is obtained from symmetry breaking, 
the little Higgs models can be product-group 
models \cite{Arkani-Hamed:2002ikv,Arkani-Hamed:2002sdy,Low:2002ws,Chang:2003zn,Chang:2003un}
or simple-group models \cite{Kaplan:2003uc,Schmaltz:2004de,Skiba:2003yf}.
A product-group model has multi-$SU(2)\times U(1)$ gauge groups that break to the electroweak 
gauge group. The most popular product-group model is the littlest Higgs model, which utilizes
the product gauge group  $[SU(2)\times U(1)]^2$~\cite{Arkani-Hamed:2002ikv}. 
 A simple-group model usually has $SU(N)\times U(1)$ gauge groups, which break to the electroweak 
gauge group. A typical simple-group model is the simplest little Higgs model, which employs 
the gauge group  $SU(3)\times U(1)$~\cite{Kaplan:2003uc,Schmaltz:2004de}. 

Obviously, these models may give slightly different Higgs couplings and thus  different 
signal rates at the LHC compared with the SM.
For example, compared with the SM predictions, at the LHC, the Higgs production and decay 
rates can be altered \cite{Han:2003wu,Cao:2008qd, Chen:2006cs,Wang:2008zw}, 
especially the di-photon signal rate always being suppressed 
\cite{Wang:2011rv} and the signal rates of the Higgs pair production being 
sizably different \cite{Wang:2007zx,Han:2009zp}. Therefore, the LHC Higgs data can constrain the little Higgs models critically \cite{Han:2013ic}. 
Another important phenomenology is in the top quark physics \cite{Cao:2006wk,Belyaev:2006jh,Han:2009zm}, 
because these models have to treat the top quark sector specially in order to cancel the quadratic 
divergence of the Higgs mass caused by the top quark loops.     

\subsection{DM and $W$-Mass in Little Higgs Models } 
The most interesting model seems to be the LHT because it can weaken the stringent constraints from the electroweak precision data and provide a DM candidate.
The littlest Higgs model, a nonlinear sigma model with a global $SU(5)$ symmetry breaking down to $SO(5)$ by a Higgs vev of order $f$,  
 predicts new heavy gauge bosons, T-quarks and a scalar particle $\Phi$, which  
cancel the one-loop quadratic divergences from the SM gauge bosons, top quarks, and Higgs self-interactions, 
respectively.   
These new particles couple to the SM particles at the tree level.
In particular, the couplings of the new heavy gauge bosons with the SM fermions 
will incur stringent constraints from the electroweak precision data,
pushing the scale $f$ of the model above a few TeV and re-incurring a little fine-tuning 
to the Higgs mass \cite{Csaki:2002qg,Hewett:2002px,Marandella:2005wd}. 
Similar to the R-parity in SUSY, a similar discrete symmetry called T-parity~\cite{Cheng:2003ju,Cheng:2004yc,Low:2004xc,Cheng:2005as} 
can be imposed  to prohibit those interactions that incur strong constraints from the electroweak 
precision data.
If the lightest T-odd particle is neutral (as with the heavy photon $A_H$) 
and the T-parity is conserved, it can serve as a DM candidate. 
For the SM down-type quarks and leptons, the Higgs
couplings of LHT have two different cases called, respectively, LHT-A and LHT-B 
\cite{Chen:2006cs,Hubisz:2004ft}. 

\begin{figure}[H]
	%\centering
	\includegraphics[width=0.7\linewidth]{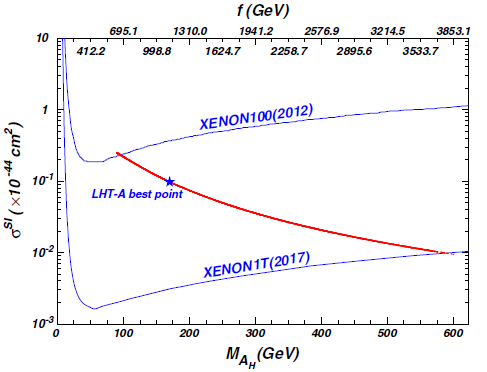} 	
	\caption{\label{LHT-DM}The LHT {parameter} %MDPI: please change hyphen to minus
 %Response: the hyphen appears in the label instead of the test.
 space allowed by 
the Planck DM relic density and the CMS Higgs data at the $2\sigma$ level, 
projected on the plane of the spin-independent scattering cross-section off the nucleon
versus the heavy photon mass. The best point is with minimal $\chi^2$ value for the 
CMS Higgs data and with the relic density closest to the measured central value.
 This figure is taken from our previous work \cite{Wang:2013yba}.
}
\end{figure}

Assuming the heavy photon $A_H$ is the lightest T-odd particle in the LHT, it can serve as a DM 
candidate. 
In \cite{Wang:2013yba,Han:2013ic}, the heavy photon relic density was found to be able to account for 
the Planck data for the small mass splitting between a mirror lepton and the heavy photon. However, the parameter space for a correct relic density was severely constrained into the range of $m_{A_H}\in [95, 600]~{\rm GeV}$ by the LHC Higgs data. As shown in Figure~\ref{LHT-DM}, under the LHC constraints, the allowed parameter space has been almost excluded by the XENON1T(2017) (we checked that the recent LZ experiment result~\cite{LZ:2022ufs} has totally excluded the parameter space).

For the little Higgs explanations of the $W$-boson mass, it was found \cite{Liu:2022dqw} 
(based on previous calculations \cite{Marandella:2005wd}) 
that the littlest Higgs model can give a sufficient contribution to explain the
CDF II measurement at the $2\sigma$ level if the scale $f$ is below 9 TeV. 
However, such a parameter space violates the T-parity of the LHT and is also not viable for the simplest little Higgs model~\cite{Liu:2022dqw}.
For the little Higgs explanations of the muon $g-2$, it is found  that the littlest Higgs model \cite{Park:2003sq,Tabbakh:2006zy} and LHT \cite{Blanke:2007db} give very small contributions to the muon $g-2$ so that the result of the FNAL plus BNL cannot be explained. 

The littlest Higgs model is found \cite{Espinosa:2004pn,Aziz:2009hk} to be unable to realize a FOPT in the allowed temperature range of the model ($0<T<4f$), assuming the UV completion factors give the SM electroweak minimum. With the same set of UV completion factors, the LHT is found \cite{Aziz:2009hk,Aziz:2010ja} to be able to realize a non-standard FOPT at the TeV scale through which a broken phase is converted into a symmetric phase.

%% fig

\section{\label{sec:4}  Two-Higgs-Doublet Extensions }
\subsection{Simplicity of 2HDMs} 
A 2HDM is an extension of the SM by merely extending the Higgs sector to 
two weak doublets of scalars, which was first proposed by T. D. Lee \cite{Lee:1973iz}.
Such an extension predicts five Higgs bosons, and in the CP-conserving version they can be classified into a neutral pseudoscalar, a pair of charged scalars,
and two neutral CP-even scalars with one being the SM-like Higgs.     
Unlike the two Higgs doublets in SUSY, which restrain the Higgs quartic interactions 
to be gauge couplings and thus predict a light Higgs boson below 135 GeV, 
a 2HDM has quite a few free couplings for the Higgs doublets and does not have the 
predictive power for a light Higgs boson. Thus, the 2HDMs cannot address 
the naturalness problem.    
The motivation of 2HDM may be that we need to extend the SM Higgs sector 
because we need more CP-violation phases for baryogenesis and we need the 
extension to provide the FOPT as discussed in the following. In other words, 
the SM Higgs sector is too simple to provide CP-violation phases and the FOPT,
both of which are needed by electroweak baryogenesis.  
Compared with the singlet extension of the SM, a 2HDM is not so simple, but it is 
much simpler than any other fancy frameworks such as SUSY or little Higgs theory. 
So, 2HDMs have simplicity and bring more light than heat.  
   
\textls[-15]{To avoid tree-level flavor-changing neutral currents, 
an additional $Z_2$ symmetry is usually imposed and hence forbids some couplings
in the Higgs potential.  
According to the $Z_2$ charge assignments of scalar doublets and the fermions,
the 2HDMs can be classified as the type-I~\cite{Haber:1978jt,Hall:1981bc}, the type-II~\cite{Haber:1978jt,Donoghue:1978cj}, the lepton-specific (or type-X), 
the flipped~\cite{Barger:1989fj,Grossman:1994jb,Akeroyd:1994ga,Akeroyd:1996he,Akeroyd:1998ui,Aoki:2009ha}, 
the inert~\cite{Deshpande:1977rw,Barbieri:2006dq,LopezHonorez:2006gr,Cao:2007rm}, etc. 
(i) In the type-I model, the imposed $Z_2$ symmetry allows one Higgs doublet 
to couple with fermions and forbids the other Higgs doublet to couple with fermions;
(ii) The type-II model is similar to the SUSY case, with one Higgs doublet coupling 
to up-type quarks while the other Higgs doublet couples to down-type quarks and leptons;
(iii)~The flipped model is same as  the type-II, except that one Higgs doublet couples  
to up-type quarks and leptons while the other Higgs doublet couples to down-type quarks;
(iv)  The lepton-specific model is rather specific, in which one Higgs doublet 
couples with quarks and the other Higgs doublet couples with leptons;
(v) In the inert model, the $Z_2$ charge is even for all the SM fields while it is odd only for 
the newly introduced Higgs doublet $\Phi_2$, which is hence called an inert doublet and has no
vev. This inert doublet cannot couple with fermions and its 
lightest neutral field is stable.         
Of course, such  $Z_2$ charge assignments seem to be ad hoc and do not make any deep sense. } 
 
Although the 2HDMs cannot address 
the naturalness problem, their phenomenology is quite rich.
Due to the multi-free parameters, the parameter space of the 2HDMs can
survive from the current LHC Higgs data and the searches for exotic scalars.
Unlike the SM, in the parameter space allowed by the current LHC Higgs data, 
the extended Higgs sector in 2HDMs can realize all three Sakharov conditions, and possibly achieve the FOPT~\cite{Jain:1993an, Dorsch:2013wja, Su:2020pjw, Basler:2016obg, Basler:2017uxn}. 
Some specific models even have the power of providing a DM candidate 
and explaining the anomalies of the muon $g-2$ and $W$-boson mass.  

\subsection{DM, Muon {$g-2$ and $W$-Mass} in 2HDMs} %Authors: we changed g-2 to $g-2$ and W to $W$ for consistency.
DM is pretty hard to explain in 2HDMs because it is not what these models were originally 
designed to solve. In a sense, this forced explanation is just like gilding the lily for the 2HDMs, 
which requires ad hoc $Z_2$ charge assignments or introducing a DM (say a singlet scalar) 
to the models.    
In the inert 2HDM, 
the lightest neutral field of the inert doublet is stable and thus can serve as 
the DM candidate \cite{Dolle:2009fn, LopezHonorez:2010eeh}. 
However, with various theoretical and experimental constraints, 
especially the DM direct detection limits, 
the parameter space for a correct relic density is highly restrained 
(for a recent study, see, e.g,~\cite{Abouabid:2023cdz}).   
A more promising scenario is the type-II 2HDM extended by introducing a real singlet 
scalar $S$ under a $Z_2$ symmetry \cite{He:2013suk}. 
In this scenario, the SM-like Higgs boson may have wrong-sign Yukawa couplings with down-type quarks, which give isospin-violating interactions between the DM and 
nucleons, relaxing the constraints from the DM direct detection \cite{Wang:2017dss}. 
Other scenarios to satisfy the  DM direct detection limits include
introducing a DM to a general 2HDM, which has blind spots for the DM 
scattering off the nucleons~\cite{Altmannshofer:2019wjb,Cabrera:2019gaq,He:2008qm,He:2011gc,He:2016mls,Chang:2017gla}, 
or introducing a DM to the lepton specific 
2HDM where the Higgs portal has suppressed couplings with the quarks \cite{Bandyopadhyay:2017tlq}. 

Putting aside the DM, which may be an axion, we check the muon $g-2$ in the 2HDMs. 
Among these models, the lepton-specific 2HDM can make sufficient contributions to the muon $g-2$ 
to explain the FNAL measurement. In this model, the Yukawa couplings of exotic Higgs bosons with
the leptons can be greatly enhanced, and the analysis in \cite{Wang:2014sda} showed 
that the muon $g-2$ explanation favors wrong-sign Yukawa couplings between the SM-like Higgs and the leptons. Because of the interference contributions between the W-loop and top quark loop, the wrong-sign Yukawa coupling of the top quark is disfavored by the 125 GeV Higgs signal strengths. However, the wrong-sign Yukawa couplings of light quarks and leptons are still consistent with the Higgs boson signal strengths.
Besides the muon $g-2$, this model is found 
to be able to provide the FOPT under the current LHC constraints \cite{Wang:2018hnw}.  
However, to explain the muon $g-2$, this model can hardly satisfy the lepton flavor universality 
in $\tau$ decays and hence some further extensions are needed, such as the lepton-specific 
inert 2HDM \cite{Han:2018znu}, the $\mu-\tau$-philic Higgs doublet model  \cite{Abe:2019bkf,Wang:2021fkn}, 
the muon-specific 2HDM \cite{Abe:2017jqo}, the perturbed lepton-specific 2HDM \cite{Crivellin:2015hha}, 
the aligned 2HDM~\cite{Ilisie:2015tra,Li:2018aov,Li:2020dbg}, the 2HDM with vectorlike 
leptons \cite{Dermisek:2022hgh}, the inert 2HDM \cite{Fan:2022dck}, etc.    
 
To jointly explain  the muon $g-2$ and the $W$-boson mass, some specific 2HDMs can make it, such as 
the 2HDM with $\mu-\tau$ LFV interactions \cite{Han:2022juu}, 
the lepton-specific 2HDM with a Higgs-phobic light pseudoscalar \cite{Kim:2022hvh}, 
the  inert 2HDM with an inert charged Higgs singlet plus a vector-like singlet quark 
and two neutral leptons \cite{Chen:2023eof}, 
the 2HDM plus an additional light pseudoscalar and a stable isosinglet massive
fermion~\cite{Arcadi:2022lpp}, etc. 
All these specific models seem to be a little unnatural or weird. In the following, we take 
the 2HDM with $\mu-\tau$ LFV interactions as an example to show the joint explanation. 
Of course, without the  muon $g-2$, the explanation of the $W$-boson mass and the FOPT can be
relatively easier in the 2HDMs, albeit sensitive to the mass splittings of the exotic Higgs bosons
\cite{Song:2022xts,Ghorbani:2022vtv}.     

\begin{figure}[H]
	%\centering
\hspace{-6pt} {}	\includegraphics[width=0.96\linewidth]{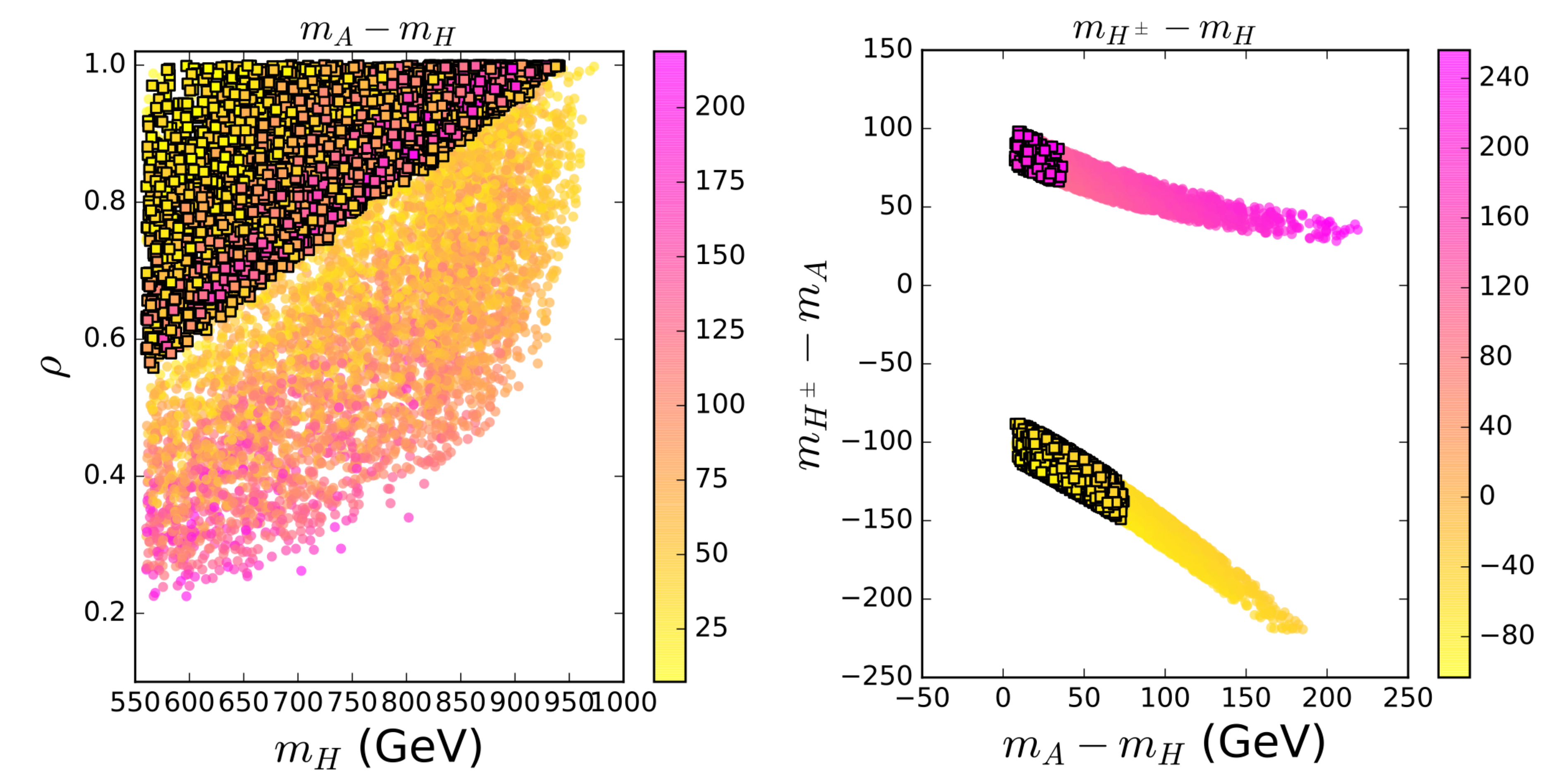} 	
	\caption{\label{2HDM-mw-gm2}Scatter plots of the parameter space of the 2HDM with $\mu-\tau$ LFV interactions: 
the dark squares (light bullets) satisfy the data of the muon $g-2$ and the $W$-boson mass at the $2\sigma$ 
level with (without) the constraints of  $\tau$-decays. 
 This figure is taken from our previous {work}
  \cite{Han:2022juu}.
}
\end{figure}

The 2HDM with $\mu-\tau$ LFV interactions is a kind of inert 2HDM except that 
a $Z_4$ symmetry is introduced and it allows the inert Higgs doublet to couple 
with $\mu-\tau$ \cite{Abe:2019bkf}.  
Only the exotic Higgs bosons from the inert doublet have  $\mu-\tau$ 
LFV Yukawa couplings, while the SM-like Higgs boson has the SM couplings with 
the gauge bosons and fermions.
The analysis showed  \cite{Han:2022juu}
that under current experimental constraints this model has some  parameter space
to simultaneously satisfy the $W$-boson mass and the muon $g-2$ as well as the lepton universality 
in $\tau$-decays. As shown in Figure~\ref{2HDM-mw-gm2}, such a parameter space is rather narrow, which requires 
tight mass splittings among the exotic Higgs bosons ($H, A, H^\pm$). Considering the joint bounds of the 125 GeV Higgs signal, the DM relic density, and the DM detection experiments, there are three allowed DM mass regions in the inert 2HDM: $m_{DM}\sim \frac{m_h}{2}$, 73 GeV $<m_{DM}<$ 75 GeV, and $m_{DM}>$ 500 GeV.

\vspace{-6pt} {}

\section{\label{sec:5}  Singlet Scalar Extensions }
\subsection{Cosmic Phase Transition in Singlet Scalar Extensions }
\begin{figure}[H]
	%\centering
	\includegraphics[width=0.56\linewidth]{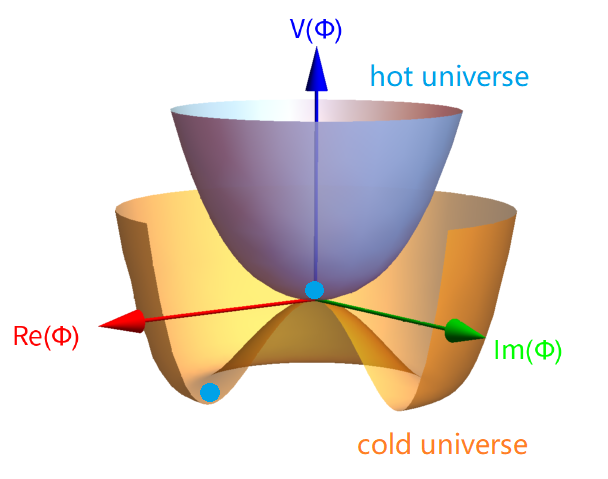} 
	\caption{\label{fig-potential}
The likely shapes of the Higgs potential at the early hot universe and the cold universe.}
\end{figure}

As shown in Figure~\ref{fig-potential}, the early hot universe may have a simple U-shape Higgs 
potential, while the cold universe may have a Mexican-hat Higgs potential. The transition property 
between the two shapes is very sensitive to the form of Higgs potential. 
If the net baryon number is generated by the electroweak baryogenesis \cite{Anderson:1991zb},  
the Higgs sector of the SM, which merely gives a rapid smooth 
cross-over \cite{Csikor:1998eu,Kajantie:1996mn} instead of a phase transition, 
must be extended to realize a strong FOPT. Such an EWPT 
occurs when the temperature of the universe decreases from an extremely high 
value to near 100 GeV, and then the universe deviates from the thermal equilibrium
to realize baryogenesis. When the phase transition is completed, the universe enters into the electroweak broken phase and the Higgs field develops a non-zero value.

\vspace{-10pt} {}
%% fig

To achieve a strong FOPT and also account for the cosmic cold DM, 
some simple singlet extensions will make it, such as the xSM~\cite{McDonald:1993ex,Cline:2013gha,Han:2016gyy,Beniwal:2017eik,Arsenault:2022xty,Palit:2023dvs}.
A slightly more complex model is the 2HDM plus a singlet (2HDM+S) \cite{He:2013suk} 
and the NMSSM \cite{Ellwanger:2009dp}. 
Note that, currently, the xSM as an explanation for DM has a very narrow 
parameter space, with the scalar DM mass being near the Higgs resonance (56--62 GeV) 
or above 1 TeV \cite{Cline:2013gha,GAMBIT:2017gge,Athron:2018ipf}, which can be relaxed by introducing some high dimensional operators \cite{Das:2020ozo}.   
Of course, the mysterious DM may just reside in the dark sector or be called the hidden sector, which interacts with the visible sector via the Higgs portal very weakly (for recent studies, see, e.g.,~\cite{Wang:2022akn,Jiang:2023xdf}).
In this case, the dark sector scalar potential may also trigger a FOPT in the early universe and 
the only way to access it is through detecting the induced gravitational wave \cite{Wang:2022akn}.  

\begin{figure}[H]

	\includegraphics[width=.95\linewidth]{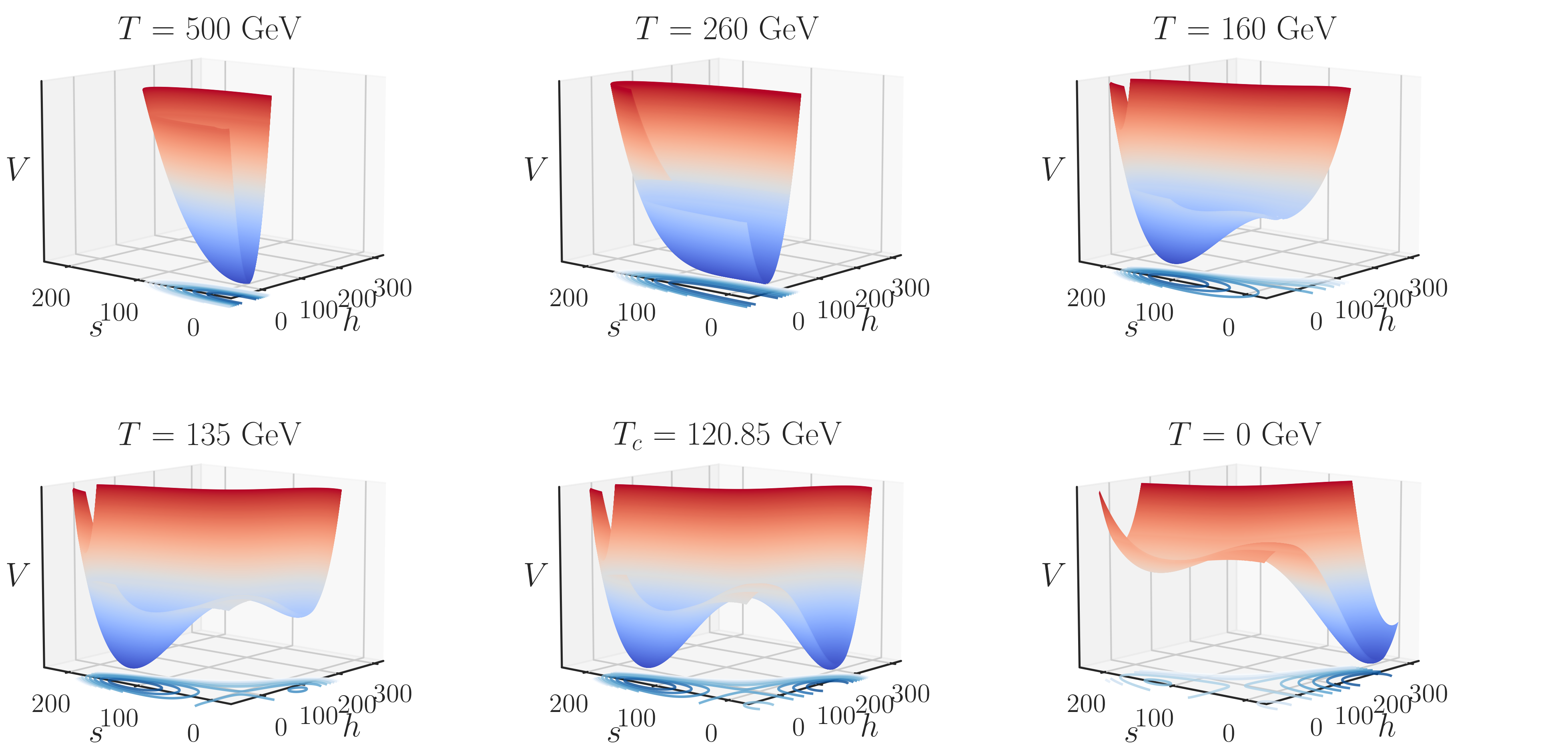} 
	\caption{\label{xSM-vacuum-1}
An illustrative diagram of effective potential developing as the temperature is dropping for the xSM, using a benchmark point taken from our previous {work}
 \cite{Balazs:2023kuk}.  
}
\end{figure}

In singlet extensions such as the xSM, we have the Higgs field $h$ and a real singlet scalar $s$,
and the phase transition occurs usually in two steps shown in Figure~\ref{xSM-vacuum-1}:
the first step is from the symmetric phase $(h,s)=(0,0)$ to the singlet-broken phase $(h,s)=(0,v_s)$ 
while the second step is from $(h,s)=(0,v_s)$ to the electroweak vacuum $(h,s)=(v_h,0)$.

\subsection{Vacuum Stability and DM in Singlet Scalar Extensions } 
As shown in Figure~\ref{xSM-vacuum-1}, in the xSM the first-step phase transition from 
the symmetric phase $(h,s)=(0,0)$ to the singlet-breaking phase $(h,s)=(0,v_s)$ 
occurs quite early at a very high temperature. Then, for the universe to be in a correct
electroweak vacuum, the second-step phase transition from  $(h,s)=(0,v_s)$ to $(h,s)=(v_h,0)$
must subsequently happen at a low temperature. 
Obviously, only checking the vacuum situation at zero-temperature cannot guarantee the 
vacuum's stability because the second-step phase transition shown in Figure~\ref{xSM-vacuum-1} 
may not happen in the thermal evolution of the universe. In other words, if we only 
examine the vacuum at zero-temperature, we usually say we have the correct vacuum if 
(i) the electroweak vacuum is the global vacuum, or (ii)  the electroweak vacuum 
is a meta-stable vacuum (its transition time to the global vacuum is longer than 
the age of the universe). 
Our recent analysis \cite{Balazs:2023kuk} showed that these two cases should be carefully checked for 
the whole thermal history. Even if  the electroweak vacuum is the global vacuum at zero-temperature,
 the second-step phase transition shown in Figure~\ref{xSM-vacuum-1} may not happen in
the thermal evolution of the universe. For the meta-stable electroweak vacuum at
zero-temperature, the universe may always reside in the singlet-breaking vacuum, which 
never transits to this electroweak vacuum.  

This unusual effect is often overlooked in studies of the vacuum stability, and the thermal history of the universe may be like this:
In the very beginning, we have an extremely hot and dense universe with electroweak symmetry. As the universe expands and the temperature drops, bubbles with broken electroweak symmetry are formed in some regions of the plasma of the universe due to fluctuations. 
% If the driving force caused by the inside-outside pressure difference of the bubbles 
%is greater than the resistance caused by the interaction of the phase-transition field 
%with the background radiation particles, the bubbles will expand outward, collide 
%and fuse, leaving the universe in a state of broken electroweak symmetry.
If the driving force of the bubble's expansion is always less than the resistance, 
then the bubbles with broken electroweak symmetry will contract, leaving the universe 
trapped in a state with unbroken electroweak symmetry. In addition, the early universe 
may have a phase transition into other vacuums (say a color-breaking vacuum) whose 
free-energy is always lower than the free-energy of the physical electroweak vacuum.

\begin{figure}[H]
%	\centering
	\includegraphics[width=0.6\linewidth]{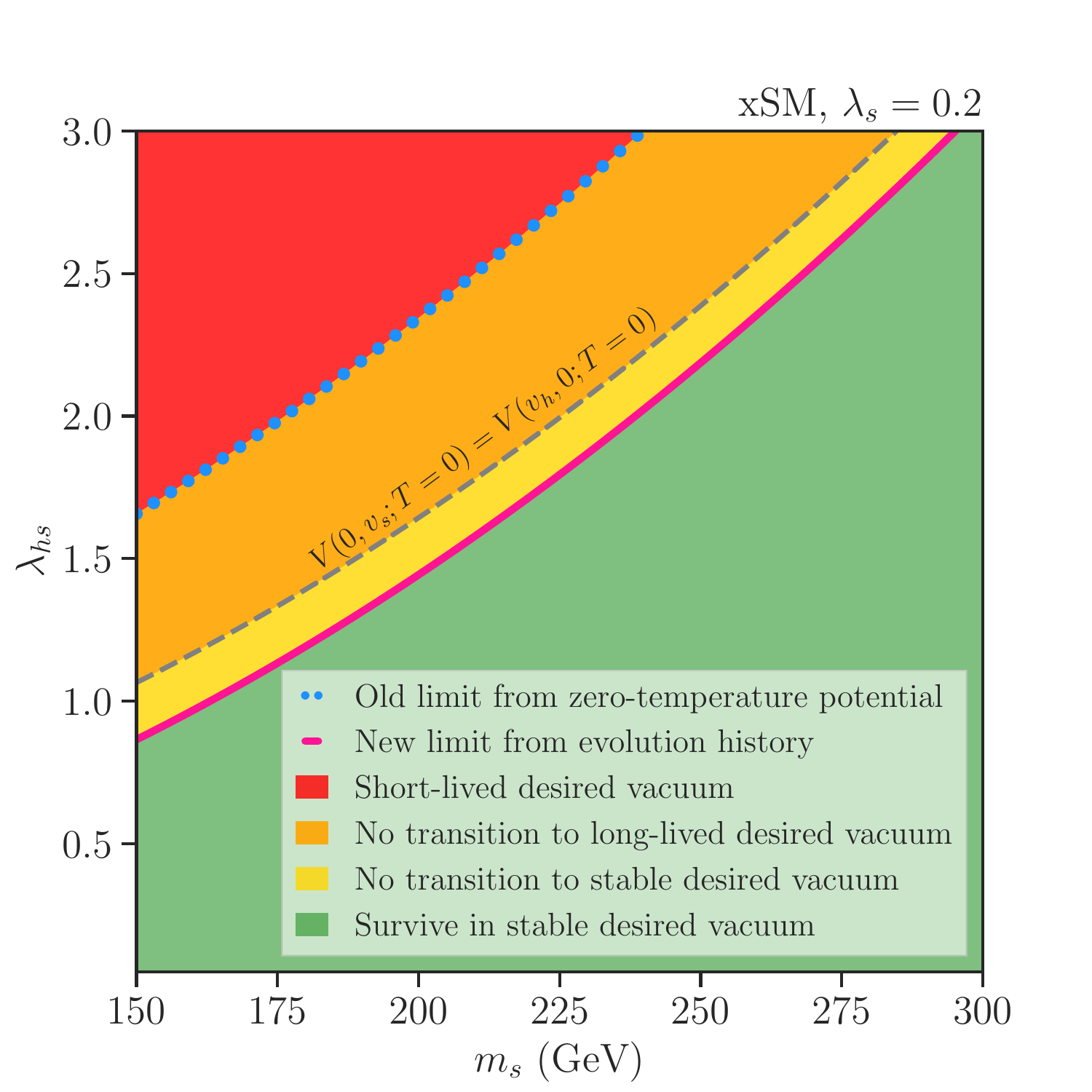} 
	\caption{\label{xSM-vacuum-2}
The xSM parameter space excluded by checking the thermal history of the universe,
taken from our previous {work}
 \cite{Balazs:2023kuk}.
}
\end{figure}

As a result, it was found \cite{Balazs:2023kuk,Kurup:2017dzf,Beniwal:2017eik,Ghorbani:2021rgs} that for the xSM a large part of the parameter space allowed by 
only checking the zero-temperature vacuum can be excluded by checking the thermal history of the universe, as shown in  Figure~\ref{xSM-vacuum-2}. 
A similar story may happen for  the phase transitions in SUSY models \cite{Cline:1999wi,Baum:2020vfl} 
or the 2HDMs \cite{Biekotter:2021ysx}. 

%% fig

\begin{figure}[H]
	%\centering
	\includegraphics[width=0.6\linewidth]{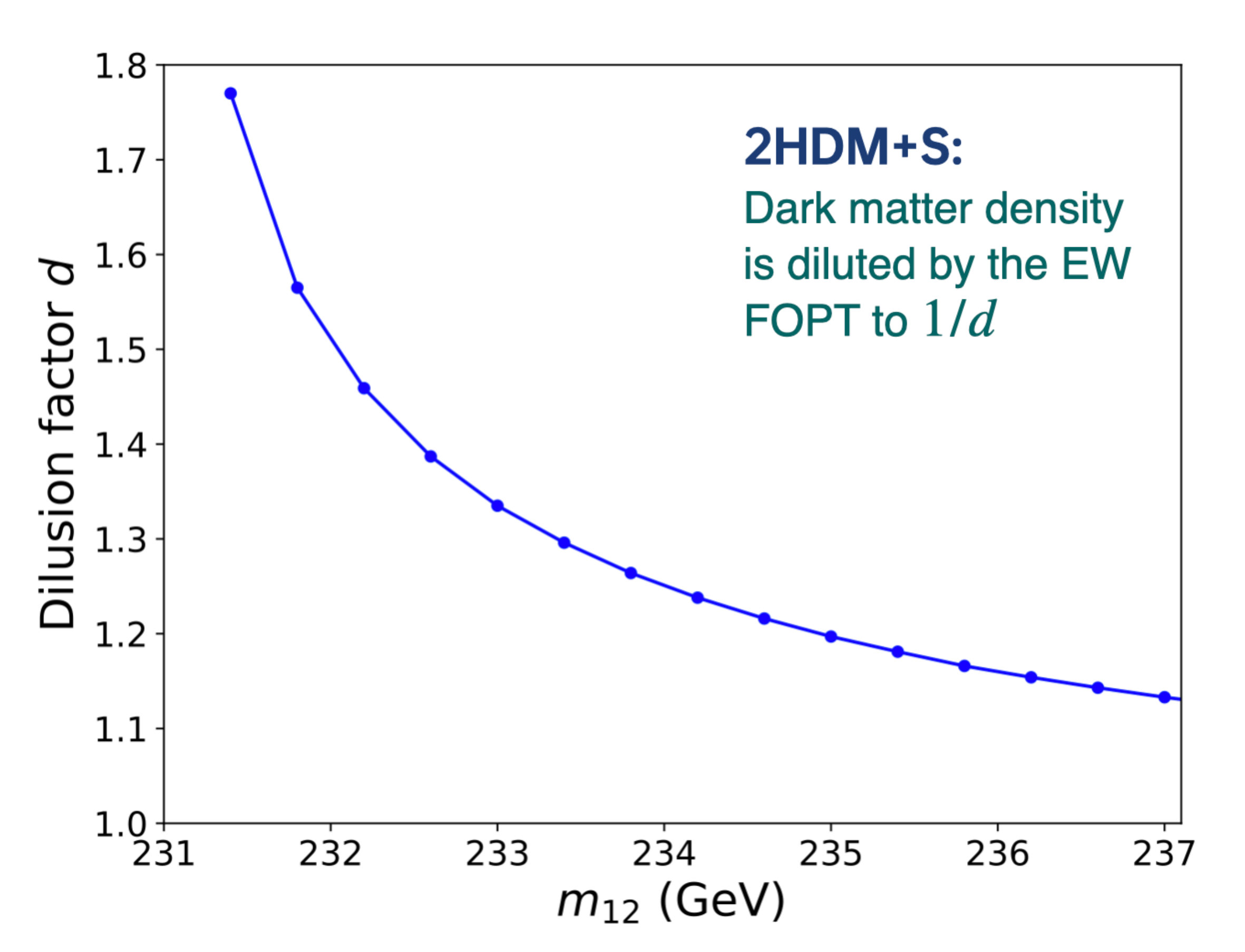} 
	\caption{\label{2HDMS-dilute}
		The dilution factor $d$ versus the mass parameter $m_{12}$ for the 2HDM+S, taken from our previous {work}
 \cite{Xiao:2022oaq}. 
		The DM density is diluted by the electroweak FOPT to $1/d$.   
	}
\end{figure}

These singlet extensions, e.g., xSM, 2HDM+S or NMSSM, can provide a cold DM candidate besides an electroweak FOPT~\cite{Ghorbani:2020xqv}. The temperature at which the  electroweak FOPT occurs may be close to the freeze-out temperature of the DM, and so the  electroweak FOPT may affect the relic density of the DM in several ways: (i)  the deviation from the thermal equilibrium caused by the phase transition may affect the size of the universe and thus change the density of DM; (ii) the particle masses may change after the phase transition, which determine the decay modes of the particles; (iii) the bubble walls formed during the phase transition may filter out most of the DM, leaving only a small amount of it. A recent analysis \cite{Xiao:2022oaq} studied the dilution of the DM relic density caused by the electroweak FOPT in the singlet extension models. It was found that the entropy released by the electroweak FOPT can maximally dilute the relic density to one third. For the xSM and NMSSM with the singlet field being relevant to the phase transition temperature, the  phase transition always happens before the DM freeze-out, and hence the dilution effect is negligible for the current relic density. However, for the 2HDM+S with the freeze-out temperature being independent of the FOPT, the dilution effect may be significant. The 2HDM+S can be regarded as the doublet extension of xSM in a sense. The Higgs state $h$ is a superposition of the neutral part of two Higgs doublets, and the electroweak FOPT is also closely related to the mixing of Higgs doublets. Therefore, as shown in Figure~\ref{2HDMS-dilute}, the electroweak FOPT can significantly dilute the DM density in the thermal history of the universe, and the dilution factor $d$ is sensitive to the doublet Higgs mass mixing term $m_{12}$\endnote{In fact, all terms in the Higgs scalar potential have an effect on the thermal history, such as the Higgs diagonal mass terms $m_{11}$ and $m_{22}$.}. 

\section{\label{sec:sum} Summary and Outlook}
For the Higgs-related BSM physics, we concisely surveyed some popular models, including
the low energy SUSY, the little Higgs theory, the 2HDMs and the simplest 
singlet extensions of the Higgs sector.
For each illustrated BSM model, either simple or complex, we see that it has its own specific features and some phenomenological power, as summarized in Table \ref{sum}. Among these models, the low energy SUSY seems to be the most compelling in phenomenology.

For the new physics models illustrated in Table~\ref{sum}, they will be directly searched for at the LHC.
The ongoing Run-III of the LHC will deliver about 150 fb$^{-1}$ of data, while the next phase of LHC, namely the HL-LHC, scheduled to start in 2027, is expected to collect 3000--4000  fb$^{-1}$ of data with
    a collision energy of 14 TeV. The parameter space of each model will be further covered substantially at the HL-LHC, e.g., the SUSY parameter space accessible at the HL-LHC is shown in Figure~10 in \cite{Wang:2022rfd}.
We also have dozens of ongoing experiments around the world looking for DM and the experiments at Fermilab and J-PARC measuring the muon $g-2$. Some colliders for precision tests, such as CEPC, FCC and ILC, are being planned, which will precisely measure the Higgs property. In addition, various gravitational wave detection experiments, such as
the LIGO, LISA, Taiji or Tianqin, will have a certain ability to explore BSM physics
related to the phase transition in the early universe.  All these experiments can allow for direct or indirect probes of these new physics models. So, leave no stone unturned.

\begin{table}[H]
\caption{\label{sum} {The} %MDPI: please confirm if keep the color in the table
%Response: we confirm to maintain the color in the table.
 phenomenological merits of some typical Higgs-related new physics models.}
\newcolumntype{C}{>{\centering\arraybackslash}X}
\begin{tabularx}{\textwidth}{lCCCCC}
\toprule
                  & \textbf{Naturalness} & \textbf{DM} & \textbf{FOPT} & \textbf{Muon \boldmath{$g-2$}} %Authors: we changed g-2 to $g-2$ for consistency.
                  &\textbf{ \boldmath{$W$}-Mass}%Authors: we changed W to $W$ for consistency. 
                  \\
\midrule
		xSM       & \textcolor{red}{\xmark}   & \textcolor{blue}{\cmark} & \textcolor{blue}{\cmark}  &  \textcolor{red}{\xmark}  &   \textcolor{red}{\xmark}   \\
		2HDMs     & \textcolor{red}{\xmark}   & \textcolor{blue}{\cmark} & \textcolor{blue}{\cmark}  & \textcolor{blue}{\cmark} & \textcolor{blue}{\cmark}   \\
        low energy SUSY        
                  & \textcolor{blue}{\cmark} \textsuperscript{1} & \textcolor{blue}{\cmark} &  \textcolor{blue}{\cmark} \textsuperscript{2}   & \textcolor{blue}{\cmark}  &  \textcolor{blue}{\cmark}   \\
  	    little Higgs theory    
                  & \textcolor{blue}{\cmark} \textsuperscript{3} & \textcolor{blue}{\cmark} &   \textcolor{blue}{\cmark}   &  \textcolor{red}{\xmark} &  \textcolor{blue}{\cmark}   \\
\bottomrule
\end{tabularx}
\noindent{\footnotesize{\textsuperscript{1} Note here that the naturalness does not mean a perfect naturalness. In SUSY, the tuning extent is at the percent level for the MSSM and at the per mille level for the CMSSM \cite{Han:2016gvr}. }}
\noindent{\footnotesize{\textsuperscript{2} The MSSM is found to be unable to realize FOPT, and here we mean the extended SUSY model such as NMSSM.  }}
\noindent{\footnotesize{\textsuperscript{3} The little Higgs theory has no quadratic divergence merely at the one-loop level.} }
\end{table}

\vspace{-12pt}

\authorcontributions{{All authors participated in discussions and writing, and contributed equally to this work. All authors have read and agreed to the published version of the manuscript.} %MDPI: For research articles with several authors, a short paragraph specifying their individual contributions must be provided. The following statements should be used ``Conceptualization, X.X. and Y.Y.; methodology, X.X.; software, X.X.; validation, X.X., Y.Y. and Z.Z.; formal analysis, X.X.; investigation, X.X.; resources, X.X.; data curation, X.X.; writing---original draft preparation, X.X.; writing---review and editing, X.X.; visualization, X.X.; supervision, X.X.; project administration, X.X.; funding acquisition, Y.Y. All authors have read and agreed to the published version of the manuscript.'', please turn to the  \href{http://img.mdpi.org/data/contributor-role-instruction.pdf}{CRediT taxonomy} for the term explanation. Authorship must be limited to those who have contributed substantially to the work~reported.
%Response: 
}
\funding{{This work} %MDPI: We moved this part to Funding section, please confirm. 
 was supported by the National Natural Science Foundation of China (NSFC) under grant Nos. 11975013, 11821505, 12075300  and 12105248,
by the Key Research Project of Henan Education Department for colleges and universities under grant number 21A140025,
by Peng-Huan-Wu Theoretical Physics Innovation Center (12047503),
by the CAS Center for Excellence in Particle Physics (CCEPP),
and by the Key Research Program of the Chinese Academy of Sciences, Grant NO.~XDPB15. %MDPI: Please add: ``This research received no external funding'' or ``This research was funded by NAME OF FUNDER grant number XXX.'' and  and ``The APC was funded by XXX''. Check carefully that the details given are accurate and use the standard spelling of funding agency names at \url{https://search.crossref.org/funding}, any errors may affect your future funding.
%Response: we have checked.
}

\dataavailability{{No new data were created in this paper.} %MDPI: We encourage all authors of articles published in MDPI journals to share their research data. In this section, please provide details regarding where data supporting reported results can be found, including links to publicly archived datasets analyzed or generated during the study. Where no new data were created, or where data is unavailable due to privacy or ethical re-strictions, a statement is still required. Suggested Data Availability Statements are available in section “MDPI Research Data Policies” at \url{https://www.mdpi.com/ethics}.
%Response: OK, we added.
} 

%\acknowledgments{{~} }%MDPI: In this section you can acknowledge any support given which is not covered by the author contribution or funding sections. This may include administrative and technical support, or donations in kind (e.g., materials used for experiments).
%Response: we don't have any other support.

\conflictsofinterest{{The authors declare no conflict of interest.} %MDPI: Declare conflicts of interest or state ``The authors declare no conflict of interest.'' Authors must identify and declare any personal circumstances or interest that may be perceived as inappropriately influencing the representation or interpretation of reported research results. Any role of the funders in the design of the study; in the collection, analyses or interpretation of data; in the writing of the manuscript; or in the decision to publish the results must be declared in this section. If there is no role, please state ``The funders had no role in the design of the study; in the collection, analyses, or interpretation of data; in the writing of the manuscript; or in the decision to publish the~results''.
%Response: OK, we added.
}

\abbreviations{Abbreviations}{ %MDPI: please confirm we change it to below format.
%Response: we have confirmed.

\noindent 
\begin{tabular}{@{}ll}
{BSM}&{beyond the standard model}\\
{LHC}&{Large Hadron Collider}\\
{SM}&{standard model}\\
{SUSY}&{supersymmetry}\\
{2HDM}&{two-Higgs-doublet model}\\
{EWPT}&{electroweak phase transition}\\
{FOPT}&{first-order phase transition}\\
{MSSM}&{minimal supersymmetric standard model}\\
{UV}&{ultraviolet}\\
{NMSSM}&{next-to-minimal supersymmetric standard model}\\
{HL-LHC}&{High-luminosity Large Hadron Collider}\\
{LSP}&{lightest supersymmetric particle}\\
{NLSP}&{next-to-lightest sparticle}\\
{WIMP}&{weakly interacted massive particle}\\
{DM}&{dark matter}\\
{GUT}&{grand unification theory}\\
{CMSSM}&{constrained minimal supersymmetric standard model}\\
{mSUGRA}&{minimal supergrivity}\\
{2HDM+S}&{2HDM plus a singlet}\\
{LHT}&{littlest Higgs model with T-parity}\\
{vev}&{vacuum expectation value}\\
{LFV}&{lepton flavor violation}\\
{xSM}&{SM plus a singlet}
\end{tabular}}

\begin{adjustwidth}{-\extralength}{0cm}
\printendnotes[custom]
%\centering %% If there is a figure in wide page, please release command \centering
\newpage
\reftitle{References}

\PublishersNote{}
\end{adjustwidth}

\end{document}